\magnification=\magstep1
\tolerance 500
\rightline{TAUP 2526-98}
\rightline{15 October, 1998}
\vskip 2 true cm
\centerline{\bf Effect of Non-Orthogonality of Residues in the Wigner-Weisskopf
model}
\centerline{\bf on}
\centerline{\bf Regeneration for the Neutral $K$ Meson system}
\centerline{Eli Cohen and L.P. Horwitz\footnote{*}{Also at Department
of Physics, Bar Ilan University, Ramat Gan, Israel}}
\centerline{Department of Physics and Astronomy}
\centerline{Raymond and Beverly Sackler Faculty of Exact Sciences}
\centerline{Tel Aviv University, Ramat Aviv 69978, Israel}
\vskip 2 true cm
\noindent {\it Abstract\/}: We  review  the application of the
Wigner-Weisskopf model to the two-channel decay problem for the
neutral $K$ meson system in the resolvent formalism.  The residues in
the pole approximation are not orthogonal, leading to additional  interference
terms in the $K_S-K_L \,\, 2\pi$ channel.  We show that these terms
lead to non-trivial changes in the exit beam in comparison to the result
 calculated with the assumptions of Lee, Oehme and Yang, and Wu and Yang,
 corresponding to semigroup evolution for which the pole residues are
 orthogonal, and hence appear to rule out the applicability of the
Wigner-Weisskopf model for the computation of regeneration in the neutral
$K$ meson system.
\vfill
\break

\par In $1957$  Lee, Oehme and Yang$^1$ constructed a generalization of the
Wigner-Weisskopf$^2$ decay model in pole approximation, in terms of a
two-by-two non-Hermitian effective Hamiltonian, leading to an exact semigroup
 law of evolution for the two channel $K^0$ decay.  Wu and Yang$^3$ developed,
from this model, an effective parametrization of the $K^0\rightarrow 2\pi$
decays, resulting in a phenomenology that has been very useful in describing
the experimental results. Since the early 1970's , Monte Carlo reproduction
of the data, establishing the parameter values, has been remarkably accurate
in the energy 60-120 Gev of the kaon beam$^3$.
  \par We wish to point out
 that the Wigner-Weisskopf model$^2$ is not, in fact, consistent with such
 semigroup evolution in pole approximation, and that the difference between
 the Wigner-Weisskopf prediction and semigroup evolution could, in
 principle, be seen experimentally.  The experimental data may
therefore
 rule out the applicability of the Wigner-Weisskopf
theory even on the level of the pole approximation.
 In this letter, we consider the structure of the two channel Wigner-Weisskopf
model on the regeneration process.We treat elsewhere the $K_S-K_L\,\, 2\pi$
interference channel at the output of a regenerator$^4$.
   \par Regeneration systems are designed to reconstitute the $K_S$ beam
 from a $K_L$ beam in the order of $K_S:K_L \cong 10^{-3}:1$, since the $K_S$
 beam decays rapidly with high branching ratio to $2\pi$, while the $K_L$
beam decays to $2\pi$ only on the order of $10^{-3}$, the measure of
$CP$  violation.
  \par The Wigner-Weisskopf model of particle decay (we first treat
the one-channel case) assumes that there is an initial state in the
Hilbert space representation of an unstable system; in the course of time,
 Hamiltonian action evolves this state, and the component that remains
in the $\phi$ direction is called the survival amplitude:
$$   A(t) = (\phi, e^{-iHt}\phi).   \eqno(1)$$
\par Defining the reduced resolvent, analytic for $z$ in the
upper half plane, in terms of the Laplace transform
$$ \eqalign{ R(z) &= -i\int_0^\infty
\,\exp^{izt}(\phi,e^{-iHt}\phi)dt \cr
&= (\phi, {1 \over {z-H}}\phi), \cr} \eqno(2)$$
we see that
$$ (\phi, e^{-iHt} \phi) = {1 \over 2 \pi i } \int_C dz\,
e^{-izt} \bigl(\phi, {1 \over z-H} \phi \bigr), \eqno(3) $$
where $C$ is a contour running from $\infty$ to $0$ slightly above the
real axis, and from $0$ to $\infty$ slightly below.  The lower contour
can be deformed to the negative imaginary axis, providing a
contribution
which is small except near the branch point, and the upper contour can
be deformed downward (with suitable conditions on the spectral
function) into the second Riemann sheet, where a resonance pole may
appear.  Assuming that the resonance contribution is large compared
to the ``background'' integrals, the reduced motion of the survival amplitude
is well-approximated by
$$ (\phi, e^{-iHt} \phi) \cong g e^{-iz_0t}, \eqno(4) $$
where $g$ is the $t$-independent part of the residue ($\approx$
unity), and
$$ z_0 = E_0 - i{\Gamma_0 \over 2}$$
is the pole position. It appears that this evolution could be
generalized for the two channel case by replacing $z_0$ by an
effective $2 \times 2$ non-Hermitian Hamiltonian.  As we shall see,
this structure is not generally admitted by the Wigner-Weisskopf
model.
\par For the two-channel system, consider a state $\phi_i, \,\,\,i=1,2$;
let us calculate the decay amplitude $\phi_i \rightarrow \vert
\lambda_j \rangle,\, j= 1,2$, the continuum accessible by means of the
dynamical evolution $e^{-iHt}$:
$$  \sum_j \int d\lambda \, \vert \langle \lambda_j \vert e^{-iHt}
\vert \phi_i) \vert^2 = 1 - \sum_j \vert (\phi_j , e^{-iHt}
\phi_i)\vert^2.  \eqno(6) $$
Here, the survival amplitude corresponds to the $2 \times 2$ matrix
$$ A_{ij}(t)= (\phi_i, e^{-iHt} \phi_j). $$
This amplitude can be approximated in the same way as in $(4)$ by
estimating
$$ ( \phi_i, e^{-iHt} \phi_j ) = {1 \over 2\pi i} \int_C
dz \, \bigl( \phi_i , {1 \over z-H} \phi_j \bigr).\eqno(7)$$
\par It is convenient to write the $2 \times 2$ matrix reduced
resolvent in the form
$$ R_{ij}(z) =   \bigl( \phi_i , {1 \over z-H} \phi_j \bigr)= \biggl({1 \over z
-
W(z)}\biggl)_{ij}, \eqno(8) $$
where $W(z)$ is a $2 \times 2$ matrix. It is almost always true (an
exception is the matrix with unity in upper right element, and all
others zero) that
$W(z)$ has the spectral decomposition
       $$ W(z) = g_1(z) w_1(z) + g_2(z) w_2(z), \eqno(9) $$
where $w_1(z)$ and $w_2(z)$ are numerical valued, and $g_1(z), g_2(z)$ are
 $2 \times 2$ matrices with the property that
$$ g_1(z) g_2(z) = 0, \eqno(10)$$
and

$$ g_1^2(z) = g_1(z), \qquad g_2^2(z) = g_2(z), \eqno(11)$$
even though $W(z)$ is not Hermitian.  These matrices are constructed
from the direct product of right and left eigenvectors of $W(z)$, and
form a complete set
$$ g_1(z) + g_2(z) = 1.\eqno(12)$$
Suppose now that the short lived $K$-decay channel has a pole at
$z_S$. The matrix $W(z_S)$ (in the second Riemann sheet)
 can be represented as
     $$ W(z_S) =  z_S g_S(z_S) + z_S' g_S'(z_S), \eqno(12) $$
where $g_S g_S' =0$, and the corresponding eigenvalues are denoted by
$z_S, z_S'$. Then the reduced resolvent, in the neighborhood of $z
\cong z_S $ has the form
$$  R_{ij}(z) \cong {1 \over (z-z_S)(1+w_1'(z_S))}g_S(z_S) + {1 \over
(z-z_S')(1+w_2'(z_S))}g_S'(z_S), \eqno(13)$$
where $w_1'(z),w_2'(z)$ are the derivatives of the eigenvalues of
$W(z)$, of order the square of the weak coupling constant (these
functions correspond to the mass shifts induced by the weak interaction). The
residues are therefore $g_S(z_S)$ and $g_S'(z_S)$ to a good
approximation.
For the long-lived component, on the other hand, the pole occurs at
$z_L$, and at this point,
$$ W(z_L) = z_L g_L(z_L) + z_L' g_L'(z_L). \eqno(14)$$
The reduced resolvent in this neighborhood is then
$$  R_{ij}(z) \cong {1 \over (z-z_L)(1+ w_1'(z_L))}g_L(z_L) + {1 \over
(z-z_L')(1+ w_2'(z_L))}g_L'(z_L). \eqno(15)$$
Since $W(z_S)$ and $W(z_L)$ correspond to the matrix-valued function
$W(z)$ evaluated at two different points, although $g_S(z_S)$ and
$g_S'(z_S)$ are orthogonal ($g_S(z_S) g_S'(z_S) = 0 $), in general,
$g_S(z_S)$ and $g_L(z_L)$ are not.
 If there were no $CP$ violation,
the matrix would be diagonal in the $K_1,K_2$ basis, and the
two distinct idempotents would be structurally orthogonal even though they
correspond to the decomposition of the matrix at two different points.
 One can estimate the product$^{4,5}$
$$ {\rm upper}\,\, {\rm left}\,\, {\rm element}\,\,\, g_S g_L
= {\rm O}(\alpha^3) =  {\rm O}(10^{-4}),
 \eqno(16)$$
in a Lee-Friedrichs type model$^6$,
where $\alpha$ is the relative amplitude $CP$ violation (independently
of the strength of the weak interaction); the other matrix elements
are of order $\alpha^4$ or $\alpha^5$. The regeneration problem was treated
 previously$^5$ with emphasis on the question of obtaining the maximal
 possible correction to the semigroup decay law$^{1,2}$.  We re-examine
 this problem in the following to establish a lower bound for such corrections,
and find that the Wigner-Weisskopf predicitons are indeed incompatible with
the experimental results.
\par We now study the composition of the beam in a
regenerator.
The usual method for the computation of regeneration is not applicable
 if the $U(t)$ do not form a semi-group.  It is generally induced by
 placing a slab of material, such as copper, in the $K$ meson beam.
\par At each scattering in the regenerator, the $K_0$ and ${\overline K}_0$
 parts of the $K_S$ and $K_L$ are altered selectively by the amplitudes
 $f$, ${\overline f}$, the scattering amplitudes for $K_0$,
 ${\overline K}_0$ on the nuclear target.\footnote{*}{G.Charpak and
M.Gourdin:CERN $67-18$.Our $f$ corresponds to $2\pi i{N\over {m}}\Delta t
f^{K^0}$, where $f^{K^0}$ is the amplitude used in this reference,
 and $\Delta t={m\over{k}}\Delta z$ is the time step of our iterative
 process ($\Delta z$ is the mean lattice spacing of the material). }
  After scattering, the new linear combination propagates freely
according to $U'(t)$.
\par In terms of $K_0$ and ${\overline K}_0$, the effect of a single
 scattering is,  $$S\vert K_0\rangle = (1+f)\vert K_0\rangle \eqno(17)$$
    $$S\vert {\overline K}_0\rangle = (1+{\overline f})\vert
 {\overline K}_0\rangle \eqno(18)$$
 where $S$ is the following term:

$$ S=(1+f)\vert K_0\rangle\langle K_0\vert +(1+{\overline f})\vert
 {\overline K}_0\rangle\langle {\overline K}_0\vert \eqno(19)$$
$U(\Delta t)$ is defined, in pole approximation, in the following way:
$$ U(\Delta t)\cong e^{-iz_S\Delta t}g_S +e^{-iz_L\Delta t}g_L \eqno(20)$$
where $g_S$, $g_L$ are the approximate residues of the reduced resolvent,
 $R'(z)$, on the second sheet.We represent these idempotents by
$$g_S=\vert K_S\rangle\langle {\tilde K}_S\vert$$
$$ g_L=\vert K_L\rangle\langle {\tilde K}_L\vert$$
where $\vert K_S\rangle$, $\vert {\tilde K}_S\rangle$ are the left and
right eigenvectors of the reduced evolution matrix (9) at $z_S$
for the eigenvalue $z_S$ and $\vert K_L\rangle$, $\vert
{\tilde K}_L\rangle$ at $z_L$ for the eigenvalue $z_L$.
Let us assume that the beam has initially the following form:
$$ \vert \psi\rangle = \alpha\vert K_0\rangle + \beta\vert
{\overline K}_0\rangle \eqno(21) $$
Then,
$$ U(\Delta t)S\vert \psi\rangle = U(\Delta t)S(\alpha\vert K_0\rangle +
\beta\vert {\overline K}_0\rangle )=$$
$$=
 \alpha(1+f)(e^{-iz_S\Delta t}g_S + e^{-iz_L\Delta t}g_L)\vert K_0\rangle +
\beta(1+{\overline f})(e^{-iz_S\Delta t}g_S + e^{-iz_L\Delta t}g_L)\vert
{\overline K}_0\rangle \eqno(22)$$
The following terms appear in this equation:
$$ \eqalign{g_S\vert K_0\rangle &= \vert K_S\rangle \langle {\tilde K}_S
\vert K_0\rangle = {1+{\tilde \epsilon}_S \over{\sqrt{2(1+\vert
{\tilde \epsilon}_S\vert ^2)}}}\vert K_S\rangle \cr
 g_S\vert {\overline K}_0\rangle &= \vert K_S\rangle \langle {\tilde K}_S
\vert {\overline K}_0 \rangle = {1-{\tilde \epsilon}_S \over
{\sqrt{2(1+\vert {\tilde \epsilon}_S\vert ^2)}}}\vert K_S\rangle \cr
 g_L\vert K_0\rangle &= \vert K_L\rangle \langle {\tilde K}_L \vert
 K_0 \rangle = {1+{\tilde \epsilon}_L \over {\sqrt{2(1+\vert
{\tilde \epsilon}_L\vert ^2)}}}\vert K_L\rangle \cr
 g_L\vert {\overline K}_0\rangle &= \vert K_L\rangle \langle
 {\tilde K}_L \vert {\overline K}_0 \rangle = {{\tilde \epsilon}_L -1 \over
 {\sqrt{2(1+\vert {\tilde \epsilon}_L\vert ^2)}}}\vert K_L\rangle \cr}
 \eqno(23) $$
Neglecting terms second order in $\epsilon$ , $f$ and changing
the basis to $\vert K_1\rangle $ , $\vert K_2\rangle $ we get that
the matrix elements of
$$U(\Delta t)S=A \eqno(24)$$     are
$$\eqalign{A_{11} &= (1+f_+)e^{-iz_S\Delta t} \cr
 A_{12} &= ({\tilde \epsilon}_S +f_-)e^{-iz_S\Delta t}+
 {\epsilon}_L e^{-iz_L\Delta t} \cr
 A_{21} &= {\epsilon}_Se^{-iz_S\Delta t} + ({\tilde \epsilon}_L +f_-)
 e^{-iz_L\Delta t} \cr
 A_{22} &= (1+f_+)e^{-iz_L\Delta t}\cr} \eqno(25)$$
where
$$\eqalign{f_+&={{f+{\overline f}}\over 2}\cr
f_-&={{f-{\overline f}}\over 2} \cr} \eqno(26)$$
$$ \eqalign{\vert K_1\rangle &= {1\over {\sqrt{2}}}(\vert K_0\rangle +
\vert {\overline K}_0\rangle )\cr
 \vert K_2\rangle &= {1\over {\sqrt{2}}}(\vert K_0\rangle - \vert
{\overline K}_0\rangle ) \cr} \eqno(27)$$
The states $\vert K_1\rangle$ and $\vert K_2\rangle$ are orthonomal:
$$\langle K_1\vert K_1\rangle = \langle K_2 \vert K_2\rangle =1$$
$$\langle K_1\vert K_2\rangle = 0$$
We define ${\epsilon}_S$, ${\tilde \epsilon}_S$, ${\epsilon}_L$, ${\tilde
\epsilon}_L$ by$^5$:
$$\eqalign{\vert K_S\rangle &= {1\over {\sqrt{1+\vert {\epsilon}_S\vert^2}}}
(\vert K_1\rangle + {\epsilon}_S\vert K_2\rangle )\cr
 \vert K_L\rangle &= {1\over {\sqrt{1+\vert {\epsilon}_L\vert^2}}}
({\epsilon}_L\vert K_1\rangle + \vert K_2\rangle)\cr
 \langle {\tilde K}_S \vert &=
{1\over {\sqrt{1+\vert {\tilde \epsilon}_S\vert^2}}}
(\langle K_1\vert + {\tilde \epsilon}_S\langle K_2\vert )\cr
 \langle {\tilde K}_L \vert &= {1\over
 {\sqrt{1+\vert {\tilde \epsilon}_L\vert^2}}}({\tilde
\epsilon}_L\langle K_1\vert + \langle K_2\vert ) \cr}\eqno(28)$$
 and,
  $$f
=e^{i\delta_0}-1$$   $${\overline f}=e^{i{\overline\delta}_0}-1$$
By multiplying the matrix $[U({\Delta t})S]$  $n$ times and keeping only first
 order in $f$,${\overline f}$, and ${\epsilon}$,
 we get the following matrix elements  for $[U({\Delta t})S]^n=A(n)$:
$$ \eqalign{A(n)_{11}&= (1+nf_+)e^{-inz_S{\Delta t}}\cr
 A(n)_{12}&= ({\tilde \epsilon}_S +f_-)e^{-inz_S{\Delta t}} \cr &+
{\epsilon}_Le^{-inz_L \Delta t} +
({\tilde \epsilon}_S
+{\epsilon}_L+f_-){{\sum}_{k=1}^{n-1}}e^{-i[kz_S+(n-k)z_L]{\Delta t}}\cr
 A(n)_{21}&= {\epsilon}_Se^{-inz_S{\Delta t}} +
 ({\tilde \epsilon}_L +f_-)e^{-inz_L \Delta t}\cr  &+
({\epsilon}_S +{\tilde \epsilon}_L +f_-){{\sum}_{k=1}^{n-1}}
e^{-i[kz_S+(n-k)z_L]{\Delta t}}\cr
 A(n)_{22}&=(1+nf_+)e^{-inz_L{\Delta t}}\cr}\eqno(29)$$
The product of the residues (16) is
$$g_Sg_L=\vert K_S\rangle\langle {\tilde K}_L\vert ({\epsilon}_L+{\tilde
\epsilon}_S)$$
$$g_Lg_S=\vert K_L\rangle\langle {\tilde K}_S\vert ({\epsilon}_L+{\tilde
\epsilon}_L)$$
and here, if these vanish,
$${\epsilon}_L+{\tilde \epsilon}_S={\epsilon}_S+{\tilde \epsilon}_L=0.$$
We obtain for this case,
$$A(n)^0_{12}=(-{\epsilon}_L+f_-)e^{-inz_S\Delta t}-{\tilde
\epsilon}_Se^{-inz_L\Delta t}+f_-{{\sum}_{k=1}^{n-1}}
e^{-i[kz_S+(n-k)z_L]{\Delta t}}$$
$$A(n)^0_{21}=-{\tilde \epsilon}_Le^{-inz_S\Delta t}+
(-{\epsilon}_S+f_-)e^{-inz_L\Delta t}+f_-{{\sum}_{k=1}^{n-1}}
e^{-i[kz_S+(n-k)z_L]{\Delta t}}.$$
 We can then substract the two cases, i.e., consider $A(n) - A(n)^0$,
 and the sums can be carried out.  We obtain
$$\eqalign{\Delta A(n)_{12}&=({\epsilon}_L+{\tilde \epsilon}_S)
(e^{-inz_L\Delta t}+{(e^{-i(n+2)(z_S-z_L)\Delta t}-
e^{-i(z_S-z_L)\Delta t})e^{-inz_L\Delta t}\over {e^{-i(z_S-z_L)\Delta t}-1}})
\cr
\Delta A(n)_{21}&=({\epsilon}_S+{\tilde \epsilon}_L)
(e^{-inz_L\Delta t}+{(e^{-i(n+2)(z_S-z_L)\Delta t}-
e^{-i(z_S-z_L)\Delta t})e^{-inz_L\Delta t}\over {e^{-i(z_S-z_L)\Delta t}-1}})
\cr }  \eqno(30)$$
where $A(n)^0_{12}$ and $A(n)^0_{21}$ are the matrices corresponding to
the orthogonal case.
For $n$ large compared to unity and we find (the $n$-dependent exponential
 factor
cancels in the ratio), for example,
  $${\Delta A(n)_{12}\over A(n)_{12}} \cong
{\epsilon_L+{\tilde  \epsilon}_S\over
({\tilde \epsilon}_S+f_-)e^{-i(z_S-z_L)\Delta t}+\epsilon_L}.$$
  The numerator is of order $10^{-4}$.  Since $(z_S - z_L) \Delta t <<1$,
 the denominator is approximately $\epsilon_L + {\tilde \epsilon}_S + f_-$.
 At $1 GeV/c$, $f_-$ is of order$^5$ $10^{-11}$, and at higher energies
 it is smaller (falling approximately as $k^{-1}$); hence
 ${\Delta A(n)_{12}\over A(n)_{12}}$ could be of order unity.  The
 validity of the the results of Monte Carlo estimates for regeneration
 according to the parametrization of ref. 2 would then appear to  rule out
 the applicability of the Wigner-Weisskopf
model for the computation of regeneration.

\bigskip
\noindent
{\it Acknowledgements}
\par We wish to thank I. Dunietz, C. Newton, B. Winstein, and T.T. Wu for
helpful discussions.
\bigskip
\frenchspacing
\noindent
{\it References}
\smallskip
\item{1.} T.D. Lee, R. Oehme and  C.N. Yang, Phys. Rev. {\bf 106}, 340
(1957).
\item{2.} T.T. Wu and C.N. Yang, Phys. Rev. Lett. {\bf 13}, 380
(1964).
\item{3.} B. Winstein, {\it et al}, {\it Results from the Neutral Kaon
Program at Fermilab's Meson Center Beamline, 1985-1997,\/}
FERMILAB-Pub-97/087-E, published on behalf of the E731, E773 and E799
Collaborations, Fermi National Accelerator Laboratory, P.O. Box 500,
Batavia, Illinois 60510.
\item{4.} E. Cohen and L.P. Horwitz, submitted for publication.
\item{5.} L.P. Horwitz and L. Mizrachi, Nuovo Cim. {\bf 21A}, 625
(1974).
\item{6.} K.O. Friedrichs, Comm. Pure Appl. Math. {\bf 1}, 361 (1948);
T.D. Lee, Phys. Rev. {\bf 95}, 1329 (1956).

\vfill
\end
\bye